\RequirePackage{fix-cm}
\documentclass[twocolumn]{svjour3}          
\smartqed  
\usepackage{graphicx}
\usepackage{physics}
\usepackage{cite}
\usepackage{xcolor}
\usepackage{enumerate}
\usepackage{bbm}
\usepackage{MnSymbol}
\usepackage{appendix}
\usepackage{setspace}
\allowdisplaybreaks

\usepackage[
  bookmarks=true,
  colorlinks,
  linkcolor=black,
  urlcolor=black,
  citecolor=black,
  plainpages=false,
  pdfpagelabels,
  final,
  breaklinks=true
]{hyperref}
\begin{document}\sloppy
\title{New schemes for creating large optical Schrödinger cat states using
strong laser fields
}

\titlerunning{New schemes for creating large optical Schrödinger cat states using
strong laser fields}        

\author{J. Rivera-Dean \and P. Stammer \and E. Pisanty \and  Th. Lamprou \and P. Tzallas \and M. Lewenstein \and M. F. Ciappina}



\institute{J. Rivera-Dean \at
ICFO - Institut de Ciencies Fotoniques, The Barcelona Institute of Science and Technology, 08860 Castelldefels (Barcelona), Spain
              \email{javier.rivera@icfo.eu}           
           \and
           P. Stammer \at
           Max Born Institute for Nonlinear Optics and Short Pulse 
			Spectroscopy, Max Born Strasse 2a, D-12489 Berlin, Germany
		    \at
           ICFO - Institut de Ciencies Fotoniques, The Barcelona Institute of Science and Technology, 08860 Castelldefels (Barcelona), Spain
           \and
           E. Pisanty \at
              Max Born Institute for Nonlinear Optics and Short Pulse 
			Spectroscopy, Max Born Strasse 2a, D-12489 Berlin, Germany
           \and
           Th. Lamprou \at
              		Foundation for Research and Technology-Hellas, Institute of Electronic Structure \& Laser, GR-70013 Heraklion (Crete), Greece \at
		Department of Physics, University of Crete, P.O. Box 2208, GR-71003 Heraklion (Crete), Greece
           \and
           P. Tzallas           \at
              		Foundation for Research and Technology-Hellas, Institute of Electronic Structure \& Laser, GR-70013 Heraklion (Crete), Greece
             \at
             ELI-ALPS, ELI-Hu Non-Profit Ltd., Dugonics t\'er 13, H-6720 Szeged, Hungary
           \and
           M. Lewenstein \at
              ICFO - Institut de Ciencies Fotoniques, The Barcelona Institute of Science and Technology, 08860 Castelldefels (Barcelona), Spain \at
              ICREA, Pg. Llu\'{\i}s Companys 23, 08010 Barcelona, Spain
              \and
                         M. F. Ciappina \at
              Physics Program, Guangdong Technion - Israel Institute
			of Technology, Shantou, Guangdong 515063, China \at
		Technion - Israel Institute of Technology, Haifa, 32000, Israel
          }

\date{Received: date / Accepted: date}

\maketitle

\begin{abstract}
Recently, using conditioning approaches on the high-harmonic generation process induced by intense laser-atom interactions, we have developed a new method for the generation of optical Schrödinger cat states \cite{Lewenstein2020}. These quantum optical states have been proven to be very manageable as, by modifying the conditions under which harmonics are generated, one can interplay between \emph{kitten} and \emph{genuine cat} states. Here, we demonstrate that this method can also be used for the development of new schemes towards the creation of optical Schrödinger cat states, consisting of the superposition of three distinct coherent states. Apart from the interest these kind of states have on their own, we additionally propose a scheme for using them towards the generation of large cat states involving the sum of two different coherent states. The quantum properties of the obtained superpositions aim to significantly increase the applicability of optical Schrödinger cat states for quantum technology and quantum information processing.
\keywords{Wigner functions \and Strong field physics \and Schrödinger cat states \and High-Harmonic Generation}
\end{abstract}

\section{Introduction}\label{intro}
In the last few years, the field of photonics has stood out as an important stage for investigations in quantum technology \cite{OBrien2009}. Due to the unique properties light has with respect to propagation and noise tolerance, quantum optical states of light are considered as one of the most valuable tools for quantum communication protocols \cite{Gisin2007}, quantum metrology \cite{Giovannetti2004} and quantum computation architectures \cite{NielsenChuangBook}. Within this direction, optical Schrödinger cat states, defined as the superposition of distinct coherent states, have been proposed as a continuous-variable candidate that could be employed in the mentioned studies \cite{Gilchrist2004}. For this reason, obtaining cat states for which the overlap among the different elements in the superposition is very small, the so-called \emph{large} cat states, has become a central problem for its employment in quantum technologies. Recently, we have demonstrated a new method based on intense laser-matter interactions, that allows for the generation of tunable optical cat states constituted by the superposition of two different coherent states (which we shall refer to hereupon as \emph{dead/alive} cat states) \cite{Lewenstein2020}. Here, we aim to take advantage of this approach in order to increase the number of distinct terms in the considered superpositions. Furthermore, we will employ the latter for generating enlarged versions of \emph{dead/alive} cat states.

At the core of our analysis lie the so-called strong-field physics processes. Strong-field physics studies light-matter interactions in high-intensity regimes, i.e. with field intensities of the order of $10^{14}$ to $10^{15}$ W/cm$^2$, accessing phenomena previously unattainable \cite{Amini2019, Krausz2009, Ciappina2017}. For instance, these processes have allowed for the generation of attosecond pulses, both in the extreme ultraviolet \cite{Drescher2001} and in the soft x-ray regimes \cite{Popmintchev2012}. Central to all this progress, from a fundamental perspective, lies the so-called three-step model \cite{Corkum1993, KulanderBook, Lewenstein1994}, which describes the interaction of the systems under consideration (typically atoms or molecules and, recently, solid materials) with the high-intensity laser pulse. In this process, the electron (1) tunnels out from the system, where it was initially bound, due to the high-intensity laser field; (2) accelerates in the continuum driven by this field; and (3) can recollide elastically or inelastically with the parent system it originated from. Particularly interesting for the context of the present paper are inelastic recollisions, which can lead to the High-Harmonic Generation (HHG) process, where a high-energy photon is generated when the electron recombines back to the ground state.

From a theoretical perspective, strong field processes have been extensively studied using a semiclassical framework \cite{Lewenstein1994}, i.e.~considering the quantum aspects of the atomic, molecular or solid-state system while keeping a classical behaviour for the laser field due to its high-photon number. Recent approaches have attempted to study these phenomena, theoretically \cite{Sundaram1990, Xu1993, Compagno1994, Gauthey1995, Becker1997, Dierstler2008, Kominis2014, Bogatskaya2016spontaneous, Bogatskaya2016polarization, Gonoskov2016, Bogatskaya2017, Bogatskaya2017spectroscopy, Magunov2017, RiveraDean2019, Gorlach2020, Yangaliev2020, Gombkoto2020} and experimentally \cite{Tsatrafyllis2017, Tsatrafyllis2019, Bloch2019}, from a fully quantum-mechanical perspective, that is, considering the quantum nature of both the laser field and the atom. However, it was in Ref.~\cite{Lewenstein2020} where we studied the depletion of the fundamental field and demonstrated, theoretically and experimentally for the first time, the quantum nature of the output light. In particular, we showed that, upon conditioning on the HHG process, we can generate optical Schrödinger cat states.

This manuscript is structured as follows: in Sec.~\ref{State:of:art} we provide a brief overview about optical Schrödinger cat states and how they can be produced. Here, we focus our attention on the novelty of the method involving HHG~\cite{Lewenstein2020}, giving a brief summary about the underlying process. After this, we review some basic concepts about the Wigner function, which constitutes a fundamental tool for the experimental characterization of non-classical states. In Sec.~\ref{Results} we present two different methods for generating superpositions of three distinct coherent states based on the initial quantum HHG states, and introduce a procedure where the former are employed for enhancing the latter. Finally, in Sec.~\ref{Conclusions}, we present some concluding comments. Our implementation is available in Ref.~\cite{FigureMaker}.

\section{State of the art}\label{State:of:art}
\subsection{Quantum-optical Schrödinger cat states}
The concept of \emph{cat states} was initially coined by E.\ Schrödinger in his famous \emph{Gendankenexperiment} \cite{Schrodinger1935} about quantum superpositions of two classically distinguishable states, exemplified by the well-known cat being in a dead and alive superposition. Within the context of quantum optics, \emph{cat states} typically refers explicitly to states which are superpositions of two distinct coherent states $\ket{\alpha_1}$ and $\ket{\alpha_2}$, i.e.
\begin{equation}\label{CatState:2}
    \ket{\psi} = 
                a_1 \ket{\alpha_1} 
                + a_2 \ket{\alpha_2},
\end{equation}
where the $a_i$ are complex coefficients satisfying normalization requirements. Besides their foundational interest, these states have proven to be very useful for practical purposes in different fields \cite{Gilchrist2004, Lvovsky2020}, such as quantum computation \cite{Ralph2003}, quantum information \cite{Sanders1992, Jeong2003,Stobinska2007} and quantum metrology \cite{Munro2002}.

The actual generation of such quantum states is considered as one of the most crucial tasks. Several methods have been proposed in the past towards this direction, with the main purpose of generating states like the one shown in Eq.~\eqref{CatState:2}, where the distance between both coherent states is sufficiently big so that their overlap is negligible. In particular, we can find techniques that condition the output of one of the ports of a beam splitter, which has been fed with a squeezed and a vacuum state, to the extraction of an odd or an even number of photons \cite{Dakna1997, Ourjoumtsev2006}. We also have methods that employ cavities in which a trapped atom can be used to change the phase of an input coherent state depending on the atomic quantum state \cite{Hacker2019}. The latest achievement in this direction is the recently developed technique which relies on conditioning measurements over high-harmonic generation processes induced by intense laser-atom interactions \cite{Lewenstein2020}. This last method constitutes the basis of the analysis presented in this manuscript.

The optical Schrödinger cat states given in Eq.~\eqref{CatState:2} can be further generalized so that they involve the superposition of more than two distinct coherent states, that is,
\begin{equation}\label{Generalized:Cat}
    \ket{\psi} = 
        \sum^{n\geq2}_{k=1} a_k \ket{\alpha_k},
\end{equation}
where $\alpha_i \neq \alpha_j \forall i\neq j$. On the one hand, states like the superposition presented above, where each $\alpha_i$ have the same amplitude but differ by a constant phase, can actually be generated by using Kerr nonlinearity and a beam splitter \cite{vanEnk2003}. On the other hand, those for which the $\alpha_i$ differ in amplitude are more complicated to create, but have proven to be useful theoretically in the context of Bell inequalities violation \cite{Wenger2003}.

\subsection{Schrödinger cat states via High Harmonic Generation}

Strong-field physics studies light-matter interactions in the regime of high-intensity electromagnetic fields, which allows us to witness highly non-linear optical phenomena such as HHG. From the theoretical perspective, and considering that a high number of photons interact with the medium under consideration, one can justifiably treat the light classically and the matter quantum mechanically \cite{Lewenstein1994}. However, by considering the quantum nature of the field, we can uncover new properties of the strong-field physics processes that are useful from a quantum optical perspective \cite{Kominis2014}. In particular, it was shown in \cite{Lewenstein2020} that by considering the quantum nature of the electromagnetic field in HHG, one is able to generate a new class of Schrödinger cat states with quantum features that depend on the HHG conditions. To show this, we consider as initial state of the whole system
\begin{equation}
    \ket{\Psi_0} = \ket{\text{g}} \otimes  \ket{\alpha} \otimes \ket{\{0\}},
\end{equation}
where the first term represents the ground state of the atomic system, the second one the coherent state of the input infrared electromagnetic field, and the last one reflects that there are no excitations in all the harmonic modes, i.e. they lie in a vacuum state. We model the dynamics of this system with the Hamiltonian
\begin{equation}
    H = H_\text{a} + H_f + V_\text{a-f}
\end{equation}
where $H_\text{a} = \hat{P}^2/2 + V(\hat{R})$ is the atomic/molecular Hamiltonian (in atomic units), $H_f$ is the field-free Hamiltonian containing all the harmonic modes up to the cutoff and $V_\text{a-f} = \hat{E}(t)\cdot \hat{R}$ describes the atom-field interaction. As we have shown previously \cite{Lewenstein2020}, after conditioning the final Schrödinger equation to the HHG process, and by considering the strong-field assumptions \cite{Lewenstein1994}, we get for the final quantum-optical state of the system \cite{Lewenstein2020}
\begin{equation}
    \ket{\psi} = \ket{\alpha + \delta \alpha} \bigotimes^\text{cutoff}_{q=2} \ket{\beta_q}.
\end{equation}

In this last expression, the input infrared field gets shifted by a displacement $\delta \alpha$ that represents the amount of photons that have been lost because of the electron-field interaction, while the harmonics appear as non-zero coherent states. In order to obtain the Schrödinger cat states, we perform a conditioning measurement that involves the generated harmonics. A clearer picture of this operation emerges from the fact that the excitation process is governed by the creation operator
\begin{equation}\label{Excitation:Mode}
    B^\dagger \propto a^\dagger e^{i\omega_L t} + \sum_{q=2}^\text{cutoff} \sqrt{q} \ b^\dagger_qe^{iq\omega_L t},
\end{equation}
where $a^\dagger$ and $b_q^\dagger$
are respectively the creation operators of the infrared mode and of its $q$th harmonic \cite{Lewenstein2020}. From Eq.~\eqref{Excitation:Mode} we see that an excitation in the fundamental mode is accompanied by excitations in the harmonic modes \cite{Gonoskov2016, Tsatrafyllis2017}. Thus, experimental analysis of the generated harmonics can be understood as a witness of the fact that a shift has been generated over the fundamental mode, and projects the final coherent state onto everything that is not $\ket{\alpha}$, i.e.~the initial state. This conditioning is carried out experimentally by means of the Quantum Tomography and Quantum Spectrometer (QT/QS) approach \cite{Lewenstein2020,Tsatrafyllis2017}. We, thus, have after such conditioning
\begin{equation}\label{Cat:HHG}
    \begin{aligned}
    P\ket{\alpha + \delta \alpha}
        &= \big(\mathbbm{1} - \dyad{\alpha}\big)\ket{\alpha + \delta \alpha}\\
        &= \ket{\alpha + \delta \alpha} 
            - \xi \ket{\alpha},
    \end{aligned}
\end{equation}
where
\begin{equation}
    \xi \equiv \braket{\alpha}{\alpha + \delta\alpha}.
\end{equation}

According to our previous definition, this kind of superpositions corresponds to an optical Schrödinger cat state. Notice that one of the greatest advantages of these strong-field cat states is that they are controllable: depending on the magnitude of $\delta \alpha$ one can move from a \emph{kitten} state to a \emph{genuine cat} state, with $\delta \alpha$ depending on the medium used for HHG and on the specific intensity of the applied laser-field. However, the appearance of $\xi$ in our equations imposes as well some limitations over the size of the generated cats: if $\delta \alpha$ is very big then $\xi \to 0$ and our cat turns into a shifted coherent state. 

\subsection{Characterizing quantum-optical states: the Wigner function}
The Wigner function was first introduced by E. Wigner in 1932 \cite{Wigner1932} within the context of quantum corrections to thermodynamic equilibrium, but it has found wide applicability in quantum optics \cite{SchleichBook, ScullyBook}. The reason for this is that it provides a natural way of characterizing quantum optical states within their quadrature representation $(x,p)$, both theoretically and experimentally \cite{Smithey1993}. It was initially formulated as a quantum analogue of the Liouville density, which gives the probability of finding a particle at a given point of the phase space. However, due to the uncertainty principle, there are some standard properties verified by the Liouville density that are not satisfied by the Wigner distribution and, for that reason, it is usually referred to as a \emph{quasiprobability distribution}.

Starting from the conventional definition ($\hbar = 1$)
\begin{equation}\label{Wigner:def}
    W(x,p) = \dfrac{1}{\pi} 
             \int^{\infty}_{-\infty} \dd y
             \mel{x + y}{\rho}{x-y}e^{-i2py},
\end{equation}
where $\rho$ is a density matrix characterizing a certain quantum state, one can prove the following properties for the Wigner function \cite{SchleichBook}:
\begin{enumerate}[(i)]
    \item it is normalized, i.e.
    \begin{equation}
        \int^{\infty}_{-\infty} \!\!\!\!\dd x
        \int^{\infty}_{-\infty} \!\!\!\!\dd p \ W(x,p) = 1;
    \end{equation}
    \item its marginals recover the probability distributions of the corresponding state along the different quadratures, i.e.
    \begin{equation}
        \begin{aligned}
        &\mathcal{P}(x) 
        = \mel{x}{\rho}{x} =\int^{\infty}_{-\infty} \!\!\!\!\dd p \ W(x,p),\\
        &\mathcal{P}(p)
        = \mel{p}{\rho}{p}
        = \int^{\infty}_{-\infty} \!\!\!\!\dd x \ W(x,p);
        \end{aligned}
    \end{equation}
    \item the overlap between two quantum states can be written in terms of their respective Wigner function representations, i.e.
    \begin{equation}
        \Tr(\rho_1 \rho_2)
            = \pi \int^{\infty}_{-\infty}\!\!\!\! \dd x
                  \int^{\infty}_{-\infty}\!\!\!\! \dd p
                        \ W_{\rho_1}(x,p)W_{\rho_2}(x,p),
    \end{equation}
    where $W_{\rho_1}$ ($W_{\rho_2}$) is the Wigner function associated to the quantum state $\rho_1$ ($\rho_2$). By setting this overlap to zero it follows that the we can find situations where the Wigner function adopts negative values, contrarily to what would happen with a well-defined probability distribution;
    \item it satisfies the following inequality
    \begin{equation}
        -\dfrac{1}{\pi} \leq W(x,p) \leq \dfrac{1}{\pi}. 
    \end{equation}
\end{enumerate}

\begin{figure}
    \centering
    \includegraphics[width =0.45\textwidth]{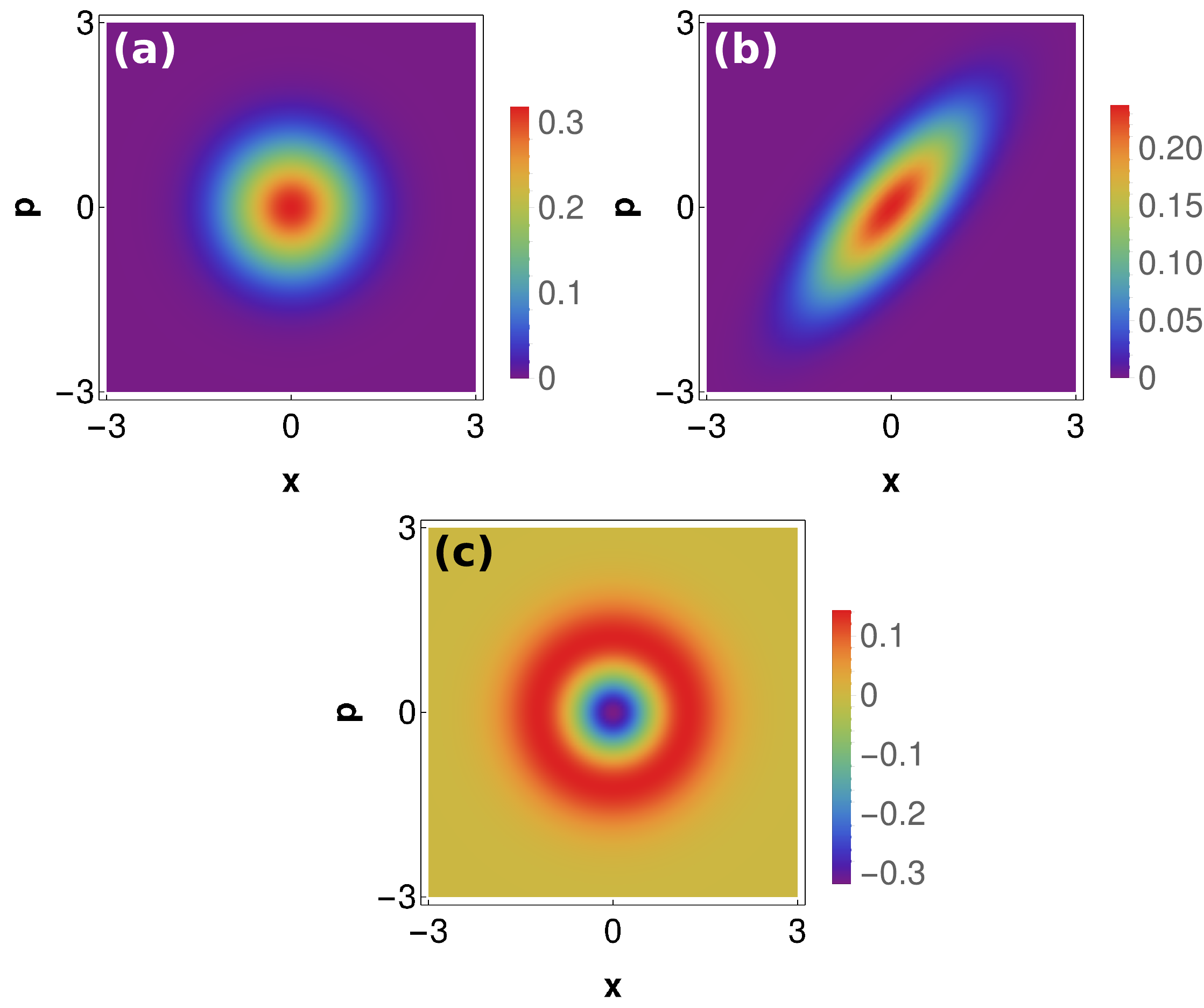}
    \caption{Wigner functions of (a) a vacuum state, (b) a squeezed state and (c) a Fock state with $n=1$. The first two cases correspond to Gaussian states examples but whose Wigner function have clearly a different behaviour, while the last one is a non-Gaussian state whose Wigner function depicts negative values.}
    \label{Fig1} 
\end{figure}

It is well established that Gaussian states depict a positive Wigner function that can be interpreted as proper probability densities \cite{Hudson1974} as happens, for instance, with coherent or squeezed states (see Fig.~\ref{Fig1} (a) and (c)). In fact, for pure states, Gaussian states are the only ones which lead to positive Wigner functions. This is not necessarily true for non-Gaussian pure states such as Fock states (see Fig.~\ref{Fig1} (c)), and may happen for non-Gaussian mixed states. Thus, one of the main uses of the Wigner function within quantum optics is to experimentally characterize non-classical states of light, a task that is typically approached using homodyne detection \cite{Smithey1993}. In this approach, one introduces in one of the two arms of an interferometer a fully characterized coherent state $\ket{\alpha}$ and, on the other arm, the state that wants to be studied. Then, by measuring both outputs of the beam splitter and varying the phase of the reference coherent state (known as the \emph{local oscillator}), one can recover the probability distribution along different points in the quadrature space of the unknown state. This information can then be used to extract the Wigner function of the state \cite{Smithey1993}.

\section{Results}\label{Results}
In this section, employing the quantum High-Harmonic Generation (QHHG) approach explained above, we present two schemes that generate superpositions involving three coherent states of different amplitude. The first consists of the use of an interferometer, where the QHHG process takes place in each arm, while the second consists of a generalization of the conditioning measurements that are performed with the QT/QS approach. Finally, we discuss a method for using the generated coherent-state superpositions for obtaining enlarged optical Schrödinger cat states \emph{à la} Eq.~\eqref{CatState:2}, i.e. \emph{dead/alive} cat states.

\subsection{Quantum High-Harmonic Generation as an optical element}
The method we present here considers the QHHG process as a modularized optical element of an experimental setup. This naturally implies that this approach could be extended to any other method that is able to generate Schrödinger cat states from a coherent state source. In that case, the QHHG stage should be substituted with the corresponding technique.

The configuration we consider here is shown in Fig.~\ref{Fig2}. It consists of two 50/50 beam splitters (BS), two quantum HHG elements (represented by QHHG), and additional paths that add a delay to one of the beams with respect to the other. From a theoretical point of view, we characterize each of the previous elements as follows:
\begin{figure}
    \centering
    \includegraphics[width = 0.45\textwidth]{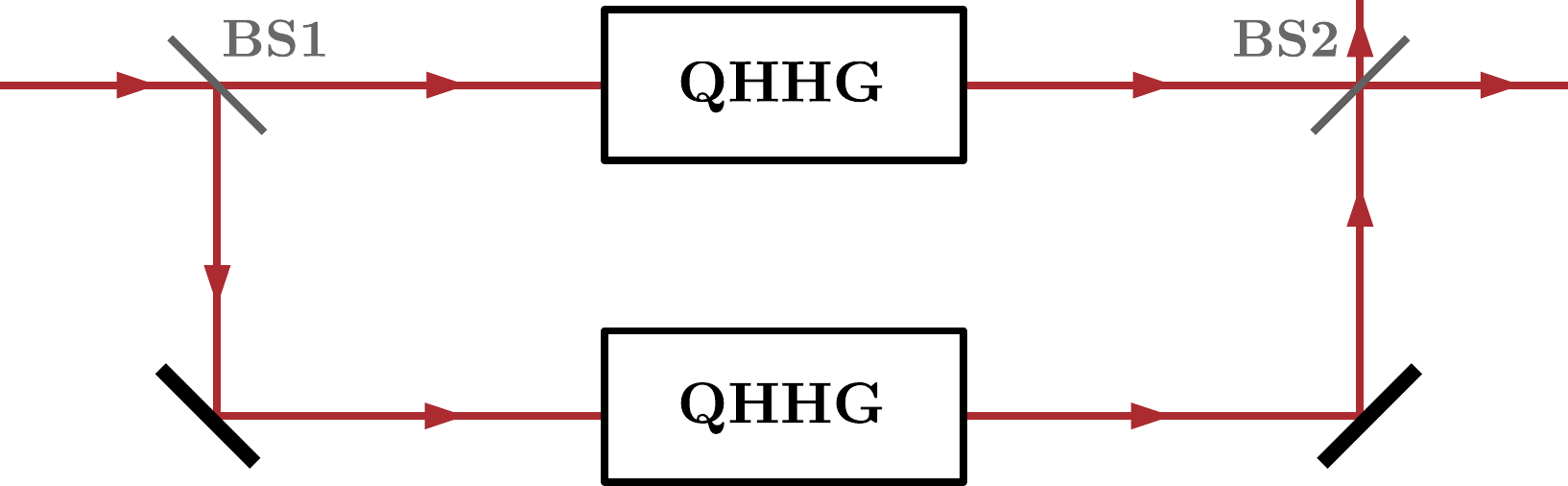}
    \caption{Optical setup consisting of two 50/50 beam splitters (BS1 and BS2), two quantum High-Harmonic Generation systems and two mirrors that redirect one of the beams. To BS1 arrives a coherent state of amplitude $\sqrt{2}\alpha$ which is used afterwards for high-harmonic generation in QHHG. After this process, we get two quantum-optical Schrödinger cat states that are mixed again in a second beam splitter (BS2).}
    \label{Fig2}
\end{figure}

\begin{itemize}
    \item \textbf{Beam splitter.} Assuming no losses, beam splitters can be described as a two-mode unitary operator that transfer energy from the input modes to the reflected and transmitted output modes. Thus, we can characterize this device as
    \begin{equation}
        B(\theta) = \exp[
                         \theta
                         (a_1a_2^\dagger 
                         - a_1^\dagger a_2)
                         ],
    \end{equation}
    where $a_1$ ($a_1^\dagger$) and $a_2$ ($a_2^\dagger$) are the annihilation (creation) operators of both input modes, and the mixing angle $\theta$ determines the ratio between reflection and transmission. In particular, if $\theta = \pi/4$, we get the so called 50/50 beam splitter. Thus, given two input coherent states $\ket{\alpha_1}\otimes\ket{\alpha_2}$ entering into the system, the output is determined by
    \begin{equation}
        \begin{aligned}
        B(\theta)\ket{\alpha_1}\ket{\alpha_2}
            &=\ket{\alpha_1 \cos(\theta) + \alpha_2 \sin(\theta)}\\
            &\hspace{0.4cm}\otimes
            \ket{-\alpha_1\sin(\theta) + \alpha_2 \cos(\theta)}.
        \end{aligned}
    \end{equation}
    
    \item \textbf{Quantum High-Harmonic Generation.} This element not only includes the medium employed for HHG, but also the experimental setup used for performing the conditioning. Therefore, according to our previous analysis, the effect of this element over a given input state $\ket{\alpha}$ is obtained from Eq.~\eqref{Cat:HHG} as
    \begin{equation}
        \text{QHHG}(\ket{\alpha})
             = \ket{\phi_\text{cat}},
    \end{equation}
    where $\ket{\phi_\text{cat}}$ is the normalized version of the state described in Eq.~\eqref{Cat:HHG}.
    
    \item \textbf{Delay path.} The objective of the delay path is to add an extra phase $\varphi$ to the corresponding field. In particular, if we start with a coherent state $\ket{\alpha}$ we get after the delay 
    \begin{equation}
        \ket{\alpha} 
            \xrightarrow[]{\text{Delay}}
                \ket{\alpha e^{i\varphi}}.
    \end{equation}
\end{itemize}

With all this set, we consider the initial state of the system to be
\begin{equation}
    \ket{\psi_0}
        = \ket{\sqrt{2} \alpha}\ket{0},
\end{equation}
where we will assume in what follows that $\alpha$ and the obtained shift $\delta \alpha$ are real quantities, positive and negative respectively. In case of zero optical path difference between the two arms of the interferometer, we obtain
\begin{equation}\label{State:BS2}
    \begin{aligned}
    \ket{\psi_\text{BS2}}
        &=
        \dfrac{1}{\sqrt{N}}
                \bigg[
                      \ket{0}\Big(
                            \ket{\sqrt{2}(\alpha + \delta \alpha)}
                            + \xi^2 
                              \ket{\sqrt{2}\alpha}
                             \Big)
                      \\
                      &\hspace{1cm}
                      - \xi
                       \Big(
                            \ket{\tfrac{\delta \alpha}{\sqrt{2}}}
                            + \ket{\tfrac{-\delta \alpha}{\sqrt{2}}}
                        \Big)
                       \ket{\sqrt{2}\alpha  +\tfrac{\delta \alpha}{\sqrt{2}}}
                \bigg],
    \end{aligned}
\end{equation}
where $N$ is a normalization constant.

Thus, we have a superposition of three distinct coherent states in both output modes of the last beam splitter. As commented before, this approach can be generalized to other setups that allow for the generation of optical Schrödinger cat states. However, the main advantage of the method employed here relies on the adaptability HHG provides over the generated cat state, through the dependence of $\delta \alpha$ on the HHG medium that is used and on the input laser intensity.
\begin{figure*}
    \centering
    \includegraphics[width = \textwidth]{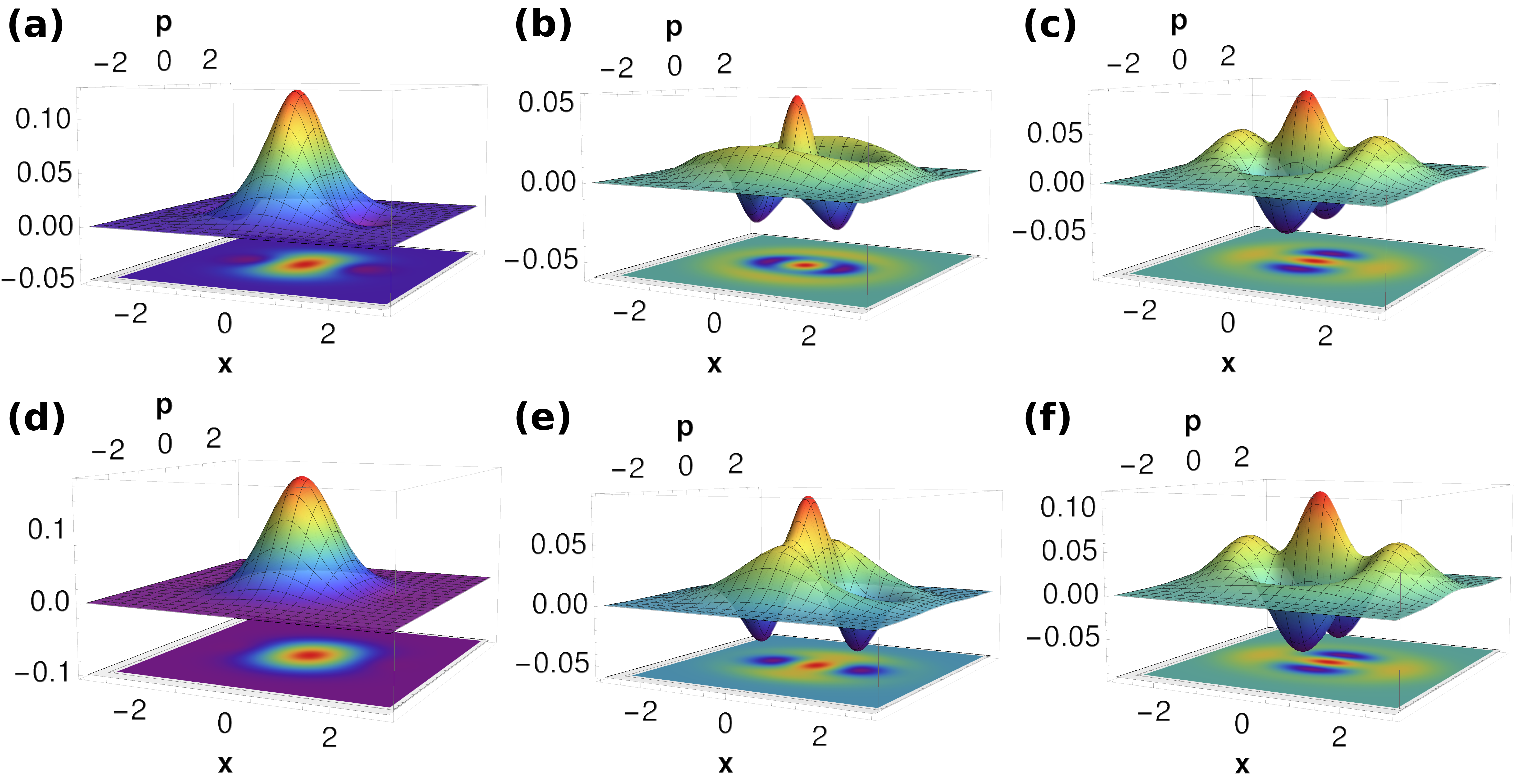}
    \caption{Wigner function for the three state superposition. Each column is defined for the different measurements defined in Eq.~\eqref{Coh:state:measu}, and are ordered accordingly to such equation from left to right. Figures (a) to (c) correspond to $\delta \Tilde{\alpha} = -0.9$, while (d) to (e) correspond to $\delta \Tilde{\alpha} = -1.3$. The measurements are strongly defined by the relative weights of the three state superposition, and in consequence this is reflected in the final form of the obtained Wigner functions.}
    \label{Fig3}
\end{figure*}
In order to recover an exact superposition of three coherent states in one of the modes, we have to constrain the output of the other mode upon a suitable measurement. Here, we will consider the situation in which such measurements are done over the second mode. In particular, we will constrain it so that we measure one of the following coherent states
\begin{equation}\label{Coh:state:measu}
    \Big\{\ket{\sqrt{2}(\alpha + \delta \alpha)},
       \ket{\sqrt{2}\big(\alpha + \tfrac{\delta \alpha}{2}\big)},
       \ket{\sqrt{2}\alpha}
    \Big\},
\end{equation}
that is, the possible coherent states that we get in the second output. In principle, this conditioning could be done over any other possible coherent state, but we restrict to these values as they are the ones we can potentially get after the beam splitter. Because of this, and from a practical point of view, we can perform this measurement by superposing the field obtained in the second output with a coherent state that has one of the previous amplitudes, but with a phase shift of half a period of the fundamental mode. Therefore, for the case in which the amplitude of both fields are equal, we will obtain a destructive interference which leads to a vacuum state. In consequence, we can understand this approach as the measurement of a vacuum state after performing a suitable displacement operation determined by one of the coherent states given in Eq.~\eqref{Coh:state:measu}.

The previous operation leads us to states in the first mode of the form
\begin{equation}\label{Pure:state}
    \ket{\psi_1}
        = \dfrac{1}{\sqrt{N_1}}
            \Big[ 
                 a \ket{0}
                 + b\ket{\delta \Tilde{\alpha}}
                 + b \ket{-\delta \Tilde{\alpha}}
            \Big],
\end{equation}
where we denote $\delta \Tilde{\alpha} = \delta \alpha/\sqrt{2}$, and $a$ and $b$ are coefficients depending on $\delta\Tilde{\alpha}$ and on the specific measurement that we are performing. Depending on the latter, the final superposition will take different forms. 

In Fig.~\ref{Fig3} we show the obtained Wigner functions for the different measurements and for two values of $\delta \Tilde{\alpha}$. In particular, figures (a) to (c) correspond to $\delta \Tilde{\alpha} = -0.9$ where the ordering, from left to right, follows the one in Eq.~\eqref{Coh:state:measu}, whereas figures (d) to (f) correspond to $\delta \Tilde{\alpha} = -1.3$. In all cases, with the exception of Fig.~\ref{Fig3} (d) where we recover a single coherent state, we can identify the same patterns for the Wigner functions: all of them are symmetric, due to the structure of the state in Eq.~\eqref{Pure:state}, and they all present negative regions that are typically located in $x = \pm \delta \Tilde{\alpha}$, unless for figures (c) and (f) where we instead have identical local maxima. 

\begin{figure}
    \centering
    \includegraphics[width = 0.48\textwidth]{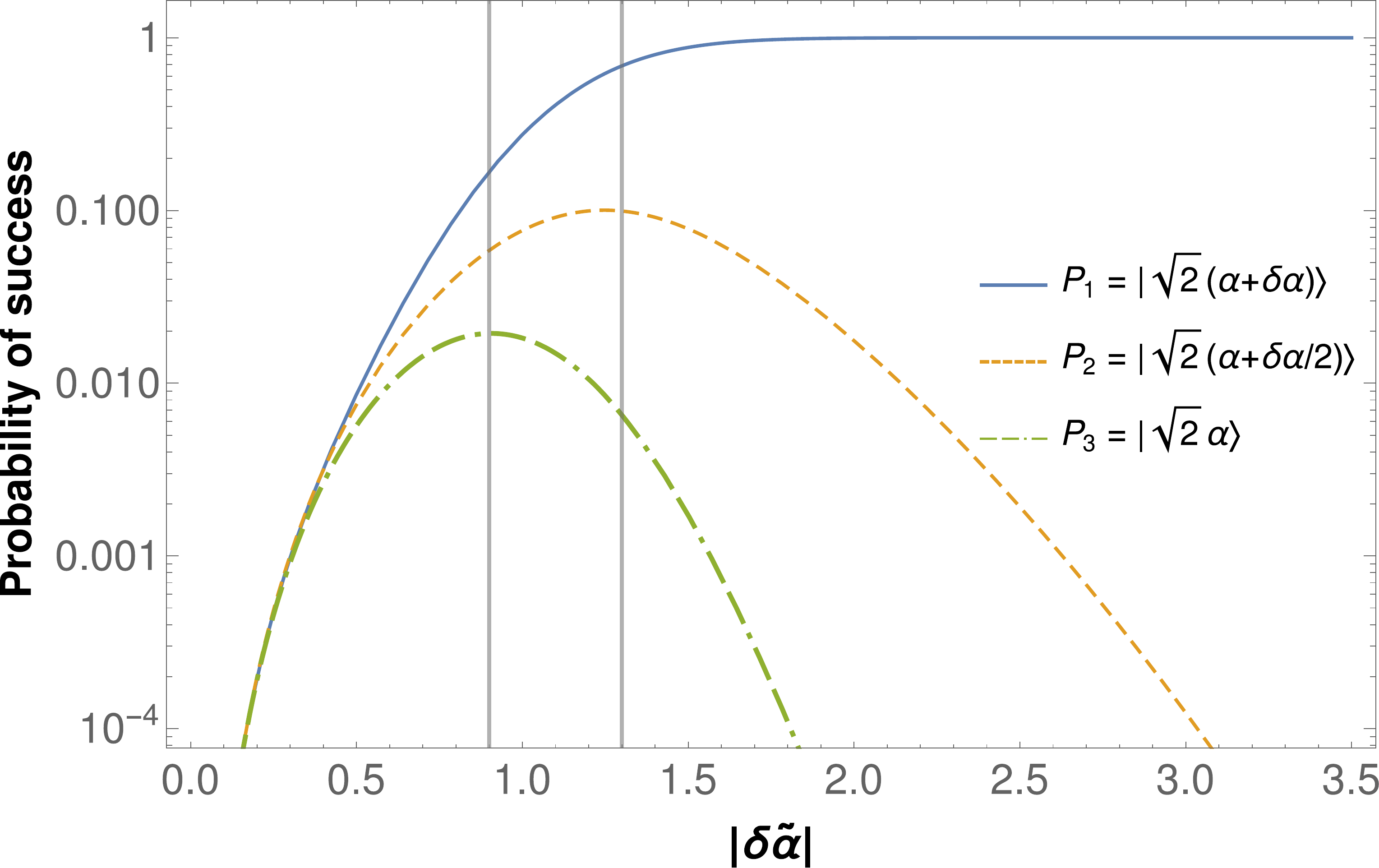}
    \caption{Probability of performing each of the measurements in Eq.~\eqref{Coh:state:measu} successfully. From left to right in the previous equation, the first measurement corresponds to the blue continuous line, the second to the orange dashed line and the third one to the green dashed-dotted line. As the value of $|\delta \Tilde{\alpha}|$ increases, the first measurement dominates while for small values it decays to zero as our three state superposition vanishes. The horizontal grey lines highlight the probabilities associated to the values of $|\delta\Tilde{\alpha}|$ used in Fig.~\ref{Fig3}.}
    \label{Fig4}
\end{figure}

The main difference between the different subplots in Fig.~\ref{Fig3} is produced by the difference in the values taken by the weights $a$ and $b$ that appear in Eq.~\eqref{Pure:state}, depending on the measurement we are performing. Thus, as we increase the value of $|\delta \Tilde{\alpha}|$, the conditioned state in the first mode tends to a coherent state for the first two measurements, in particular to a vacuum state which is the term dominating the superposition. Within this regime, the Wigner function corresponds to a Gaussian centered in $x=p=0$. For the second measurement this is a process that takes place more slowly, as for $\delta \Tilde{\alpha} = -1.3$ we still get negative regions for the Wigner function arising from the overlap between the different elements in the superposition, contrarily to what happens with the first measurement (Figs.~\ref{Fig3} (e) and (d) respectively). On the other hand, for the third measurement there are no changes at all in the final superposition, which can be seen from the fact that its Wigner function remains unperturbed upon changes in $\delta \Tilde{\alpha}$. This implies that weights between the vacuum state and the rest of the superposition in Eq.~\eqref{Pure:state} are almost equal. Moreover, one can check that, in fact, is the $b$ coefficient in the previous equation the dominating one, and in the limit of large values of $|\delta \Tilde{\alpha}|$ the Wigner function coincides with the one coming from the symmetric superposition between $\ket{\delta \Tilde{\alpha}}$ and $\ket{-\delta\Tilde{\alpha}}$. However, in order to obtain such states one has to perform the adequate measurement successfully, and this is highly determined by which state of Eq.~\eqref{Coh:state:measu} are we using. In Fig.~\ref{Fig4} we show the success probability for each of the possible measurements as a function of $\delta \Tilde{\alpha}$. As the latter becomes bigger, only the first measurement can be performed successfully, which leads to a vacuum state as mentioned previously.

\subsection{Generalization of the conditioning approach}
One of the main problems regarding the previous approach is that, due to the use of the second beam splitter (BS2), the distance between the terms in the superpositions that we can reach are relatively small, as the initial $\delta \alpha$ gets affected by a factor of $1/\sqrt{2}$. In top of this, as we have shown in Fig.~\ref{Fig4} for values of $|\delta \Tilde{\alpha}| > 1.3$, the measurement performed with the first state in Eq.~\eqref{Coh:state:measu} are more likely and, in consequence, the vacuum term of Eq.~\eqref{Pure:state} dominates the superposition, leading to a final Gaussian state. Thus, for this kind of superpositions to be useful for practical purposes such as in \cite{Wenger2003}, it seems instrumental to develop techniques that allow us to enlarge the distance between the different coherent states in the superposition, while keeping a good ratio between their probability amplitudes.

\begin{figure}[h]
    \centering
    \includegraphics[width = 0.45\textwidth]{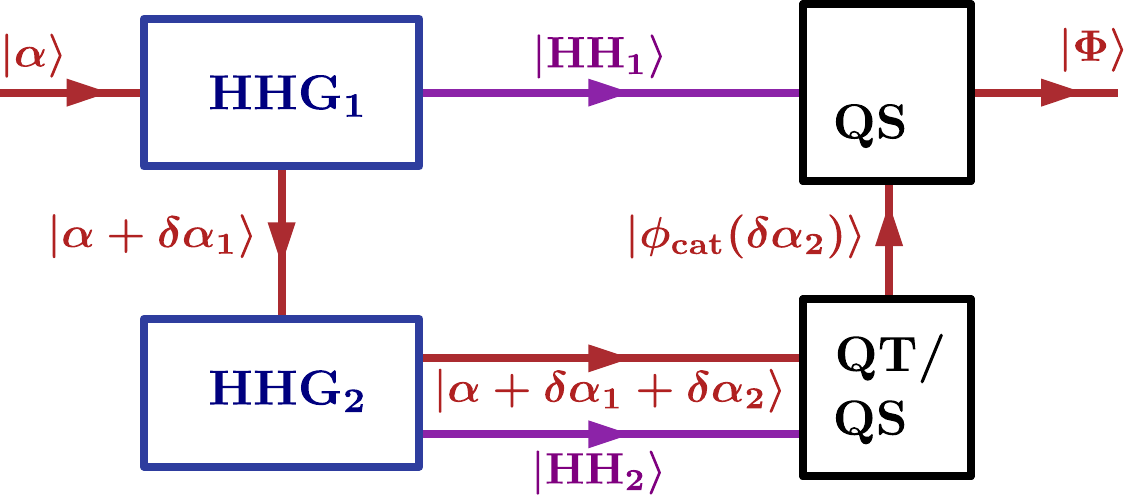}
    \caption{Experimental setup for the generalized approach. The state $\ket{\alpha}$ undergoes two HHG processes in the media HHG$_1$ and HHG$_2$. Later on, the harmonics $\ket{\text{HH}_i}$ generated in each of these interactions are used for performing the conditioning measurements with the Quantum Tomography and Quantum Spectrometer approach, which we refer to with the QT/QS box. As shown in the text, with this set-up we can naturally generate superpositions of three distinct coherent states.}
    \label{Fig5}
\end{figure}

The approach we consider here is based on the architecture shown in Fig.~\ref{Fig5}. In this case, we have several HHG processes taking place but, unlike the set-up of Fig.~\ref{Fig2}, all of them are mediated with the same mode. Therefore, each of the harmonics that are generated will define projectors with respect to different initial coherent states once the conditioning measurement is performed. Depending on the number of HHG processes considered in this scheme, we can naturally generate superpositions containing more than two coherent states without involving any beam splitter, and thus without introducing a factor $1/\sqrt{2}$ affecting their final intensity.

Let us illustrate this idea with the configuration shown in Fig.~\ref{Fig5}. In this case we start with a coherent state $\ket{\alpha}$ which undergoes HHG in a given medium, represented as HHG$_1$. We will denote the generated shift as $\delta \alpha_1$ and use $\ket{\text{HH}_1}$ as a shorthand notation for the harmonic modes generated in this system. Hence, the process taking place in this first medium is
\begin{equation}
    \ket{\alpha} \otimes \ket{\{0\}}
    \rightarrow
    \ket{\alpha+\delta \alpha_1} \otimes \ket{\text{HH}_1}.
\end{equation}

Afterwards, instead of performing the conditioning measurement (represented in Fig.~\ref{Fig5} with the QT/QS box), we use the fundamental shifted mode $\ket{\alpha + \delta \alpha_1}$ in another medium HHG$_2$. Denoting with $\delta \alpha_2$ and $\ket{\text{HH}_2}$ the generated shift and the harmonics respectively, we get in this case
\begin{equation}
    \begin{aligned}
    &\ket{\alpha + \delta\alpha_1}
    \otimes \ket{\text{HH}_1}\otimes \ket{\{0\}}\\
    &\hspace{1.5cm}
    \to
    \ket{\alpha+\delta \alpha_1 + \delta\alpha_2} 
    \otimes \ket{\text{HH}_1}
    \otimes \ket{\text{HH}_2}.
    \end{aligned}
\end{equation}

Note that the harmonics $\ket{\text{HH}_1}$ and $\ket{\text{HH}_2}$ that we have generated thus far belong to two different modes, as they are spatially separated. Moreover, both harmonic modes are uncorrelated to each other, but correlated with the fundamental mode via the creation operators $B^\dagger_1$ and $B^\dagger_2$ given in Eq.~\eqref{Excitation:Mode}, where the difference between them is that they are defined with respect to two different creation operators $b^\dagger_{q,i}$, $i = 1, 2$, affecting the $q$th harmonic in each of the space-separated modes. This implies that if we now use the QT/QS approach to perform a conditioning measurement with respect to $\ket{\text{HH}_2}$, we should consider for the definition of the corresponding projector $P_2$ the initial state that was used for its generation. More explicitly, we have
\begin{equation}
    \begin{aligned}
    \ket{\phi_\text{cat}^{(2)}} &= P_2 \ket{\alpha + \delta \alpha_1 + \delta \alpha_2}\\
        &= 
        (1-\dyad{\alpha + \delta\alpha_1})\ket{\alpha + \delta \alpha_1 + \delta \alpha_2}\\
        &=
        \ket{\alpha + \delta \alpha_1 + \delta \alpha_2}
        - \xi_2 \ket{\alpha + \delta \alpha_1},
    \end{aligned}
\end{equation}
where $\xi_2 = \braket{\alpha+\delta\alpha_1}{\alpha+\delta\alpha_1+\delta\alpha_2}$.

Now, we use the harmonics generated in the first medium, i.e. $\ket{\text{HH}_1}$, for performing a second conditioning. Similarly to what we had before, in this case the conditioning projector $P_1$ refers to the initial state used for the generation of such harmonics, that is, $\ket{\alpha}$. We get after that operation
\begin{equation}
    \begin{aligned}
    \ket{\Phi} 
    &= P_1 \ket{\phi_\text{cat}^{(2)}}
    = (1-\dyad{\alpha})
    \ket{\phi_\text{cat}^{(2)}}\\
    &= \ket{\alpha + \delta \alpha_1 + \delta \alpha_2}
        - \xi_2 \ket{\alpha + \delta \alpha_1}
        - \xi_1 \ket{\alpha},
    \end{aligned}
\end{equation}
where $\xi_1 =  \braket{\alpha}{\alpha + \delta \alpha_1 + \delta \alpha_2} - \xi_2 \braket{\alpha}{\alpha + \delta\alpha_1}$. As we see, this method provides us with a very natural way of generating the desired superposition of three coherent states.

However, one of the main drawbacks with respect to this configuration is that the $\xi_1$ constant might be very small. To avoid this, we implement between the $P_1$ and $P_2$ conditioning an amplification operation over the cat state $\ket{\phi_\text{cat}^{(2)}}$, described by the photonic displacement operator $D(\gamma)$ \cite{ScullyBook}. Thus, if we set $\delta \alpha_1 = \delta \alpha_2 \equiv \delta \alpha$ for the sake of simplicity, and consider $\gamma = - 3 \delta \alpha$, we get after the amplification process
\begin{equation}
    \ket{\Tilde{\phi}_\text{cat}^{(2)}}
    = D(-3\delta \alpha) \ket{\phi_\text{cat}^{(2)}}
    = \ket{\alpha - \delta \alpha}
    - \xi_2\ket{\alpha - 2 \delta \alpha},
\end{equation}
and after the conditioning
\begin{equation}\label{Enhanced:Superposition}
    \ket{\Phi}
        = 
            - \xi'_1 \ket{\alpha}
                +\ket{\alpha - \delta \alpha}
                - \xi_2 \ket{\alpha - 2\delta \alpha},
\end{equation}
where $\xi'_1 = \xi_2(1 - \braket{\alpha}{\alpha-2\delta\alpha})$. We can see that as $\delta \alpha$ increases, $\xi'_1 \to \xi_2$ and therefore the obtained state is symmetric around $\ket{\alpha - \delta \alpha}$. Nevertheless, one should be careful in the limit where $\delta \alpha$ is very large because then $\xi_1',\xi_2 \to 0$ and we get only a coherent state. 

\begin{figure}
    \centering
    \includegraphics[width =0.48\textwidth]{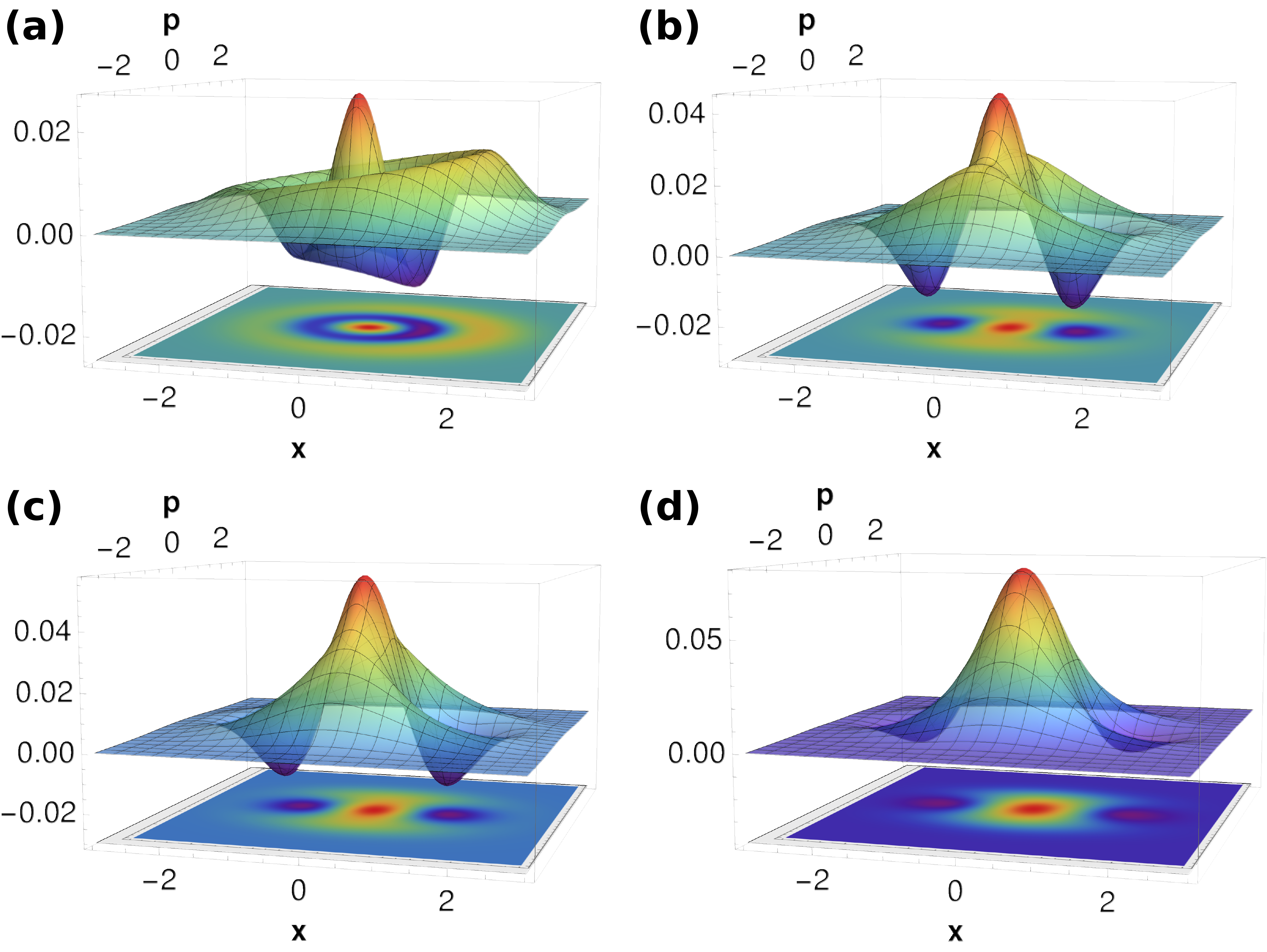}
    \caption{Wigner functions for the state obtained by the generalized measurement approach shown in Fig.~\ref{Fig5}, for (a) $\delta \alpha = -0.9$, (b) $\delta \alpha = -1.3$, (c) $\delta \alpha = -1.5$ and (d) $\delta \alpha = -2.0$. As we can see, with this method we can achieve negativities of the Wigner function for bigger values of $\delta \alpha$ compared to most successful measurement of the previously described method. Furthermore, the symmetry of the Wigner functions increases for bigger values of $\delta \alpha$. In these plots we have set $\alpha = 0$ so that the functions are centered in the origin.}
    \label{Fig6}
\end{figure}

The main distinction between Eq.~\eqref{Pure:state} and Eq.~\eqref{Enhanced:Superposition} lies in the coefficients that go along with the different coherent states in the sum, which now are only determined by the value of $\delta \alpha$, and on the distance between the three states in the superposition. While in the former the distance between the two outermost coherent states is given by $\sqrt{2} \delta \alpha$, in the latter is given by $2\delta \alpha$. The corresponding Wigner functions obtained for this state are shown in Fig.~\ref{Fig6} for values of $\delta \alpha = -0.9, -1.3, -1.5$ and $-2.0$, respectively, from (a) to (d). As we can see, the obtained Wigner functions bear some similarity with the ones obtained in Figs.~\ref{Fig3} (b) and (e), specially Figs.~\ref{Fig6} (a) and (b). However, one of the most important differences with respect to Eq.~\eqref{Pure:state} is that the obtained state is not completely symmetric, specially for small values of $|\delta \alpha|$ for which we have $\xi_1' \neq \xi_2$, as it can be clearly seen in Fig.~\ref{Fig6} (a) where the negative regions of the Wigner function located in the positive part of the $x$ axis are slightly bigger than the ones obtained in the negative side. As mentioned previously, for increasing values of $|\delta \alpha|$ the obtained Wigner function becomes more symmetric since $\xi_1' \to \xi_2$. This is clearly seen in Figs.~\ref{Fig6} (c) and (d), where there is almost no difference between the negative regions of both Wigner functions. It is remarkable as well that for values of $\delta \alpha = -2.0$ we still get distinguishable negative regions for the Wigner function, which motivates the use of this kind of three coherent state superpositions to generate large \emph{dead/alive} cat states. However, and above that value, the central coherent state becomes more important and the Wigner function tends to a Gaussian, as seen in Fig.~\ref{Fig6} (d).

\subsection{Enlarged cat states}
The two methods we have presented so far allow us to generate a superposition of three coherent states. While in the first method the distance between the two outermost states in the superposition is $\sqrt{2}\delta \alpha$, with the second approach we are able to increase this distance by a factor of $\sqrt{2}$. Given that the superposition of two well-separated coherent states, i.e. large \emph{dead/alive} cat states, are essential for practical uses \cite{Gilchrist2004}, it is natural to ask ourselves how can we use these three-state superpositions to generate enlarged \emph{dead/alive} cat states.

Inspired by \cite{Laghout2013, Sychev2017}, a straightforward way of achieving this objective is by means of a beam splitter and a conditioning measurement in one of the output modes. Given that the first method for obtaining the three coherent state superposition already contains a beam splitter, the final enhancement will not be excessively big. Thus, in this section we will focus mainly on the states obtained with the generalized measurement method as it does not contain any extra beam splitter in its definition. 

Starting from the state shown in Eq.~\eqref{Enhanced:Superposition} and postprocessing it with a 50/50 beam splitter whose other input is fed with $\ket{\alpha + \delta \alpha}$, we get
\begin{equation}
    \begin{aligned}
    \ket{\Phi_\text{BS}}
        &= \dfrac{1}{\sqrt{N}}
            \Big[
                 -\xi'_1 \ket{\Tilde{\alpha} + \tfrac{\delta\alpha}{\sqrt{2}}}\ket{-\tfrac{\delta \alpha}{\sqrt{2}}}
                 + \ket{\Tilde{\alpha}}
                    \ket{0}\\
                &\hspace{1.3cm}- \xi_2 \ket{\Tilde{\alpha} - \tfrac{\delta\alpha}{\sqrt{2}}}\ket{\tfrac{\delta \alpha}{\sqrt{2}}}
            \Big],
    \end{aligned}
\end{equation}
where $\Tilde{\alpha} = \sqrt{2}(\alpha - \delta \alpha)$ and $N$ is a normalization constant.
\begin{figure}
    \centering
    \includegraphics[width =0.48\textwidth]{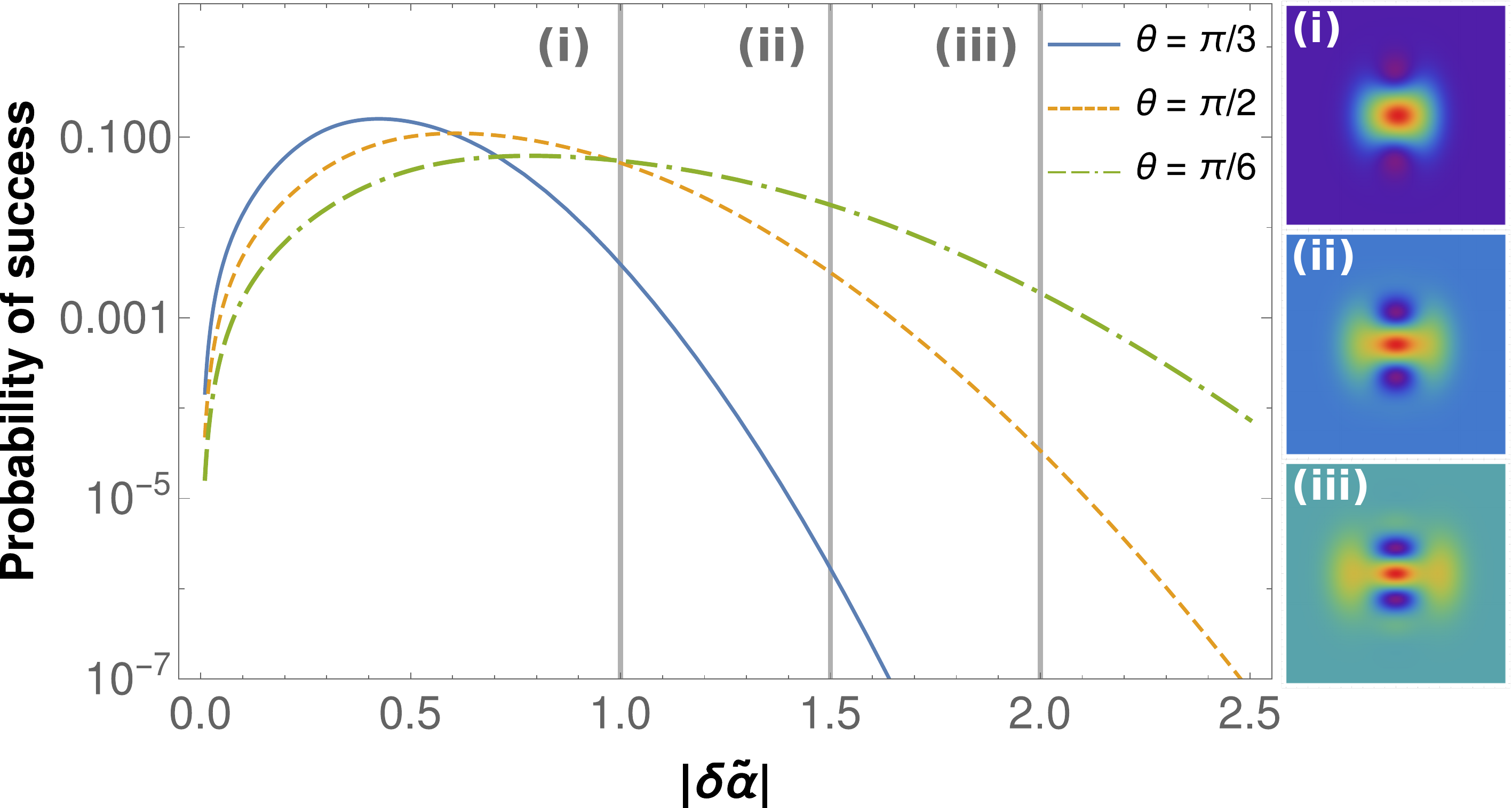}
    \caption{Probability of success when conditioning the results of the second mode to obtain an even number of photons, versus $\delta \Tilde{\alpha} \equiv \delta \alpha \cos(\theta)$ (in absolute value). We present the results for three different beam splitters characterized by mixing angles $\theta = \pi/3$ (blue continuous line), $\theta = \pi/2$ (orange dashed line) and $\theta = \pi/6$ (green dashed-dotted line). While the latter gives smaller probabilities, the decay for large values of $|\delta \Tilde{\alpha}|$ takes place more slowly which allow us to obtain bigger \emph{dead/alive} cat states. The vertical continuous grey lines correspond to the values of $|\delta \Tilde{\alpha}|$ for which we get the different Wigner functions shown in the the insets (i)-(iii) on the right. In these plots we have set $\Tilde{\alpha} = 0$ so that the functions are centered in the origin.}
    \label{Fig7}
\end{figure}

If we now condition the second output to the measurement of an even number of photons \cite{Lvovsky2020,Gerrits2010}, we obtain
\begin{equation}\label{Number:photon:state}
    \begin{aligned}
    \ket{\Phi_\text{even}}
        &= \Big( 
                \mathbbm{1} \otimes \sum_{n=1}^\infty \dyad{2n}
            \Big)
            \ket{\Phi_\text{BS}}
        \\
        &= \ket{\phi_\text{even}(\tfrac{\pi}{4})}
            \otimes
            e^{-\tfrac{|\delta \Tilde{\alpha}|^2}{2}}
            \sum^{\infty}_{n=1} \dfrac{(\delta \Tilde{\alpha})^{2n}}{\sqrt{2n!}}
            \ket{2n},
    \end{aligned}
\end{equation}
where
\begin{equation}
    \ket{\phi_\text{even}(\tfrac{\pi}{4})} =
    \dfrac{1}{\sqrt{N}}
            \Big[
                 \xi_1'\ket{\Tilde{\alpha} + \tfrac{\delta\alpha}{\sqrt{2}}}
                 + \xi_2 \ket{\Tilde{\alpha} - \tfrac{\delta\alpha}{\sqrt{2}}}
            \Big]
\end{equation}
is the non-normalized optical Schrödinger cat state that we get for the first mode. 

The state in Eq.~\eqref{Number:photon:state} is a separable state where the first mode has been projected into a \emph{dead/alive} cat state and the distance between both coherent states is $\sqrt{2}\delta \alpha$, i.e. a factor $\sqrt{2}$ bigger than the distance between the two outermost coherent states in the inputted three state superposition. Because the state in Eq.~\eqref{Number:photon:state} is separable, the partial trace with respect to the second mode naturally leads to a pure state.

One of the main advantages of this technique in comparison with the methods presented in \cite{Laghout2013, Sychev2017} is that, with this approach, we can get larger enhancements by suitably changing the beam splitter transmittance and the coherent state entering one of its inputs, provided that the three coherent state superposition is inputted in the other. This contrasts with the previously cited methods since, in those cases, we are forced to use a 50/50 beam splitter, because the two inputs are given by identical \emph{dead/alive} cat states. More specifically, for a beam splitter characterized by the mixing angle $\theta$ and a coherent state of the form $\ket{(\alpha + \delta \alpha) \tan(\theta)}$, one can show (see Appendix \ref{Appendix:A}) that the enhanced and not normalized \emph{even} optical Schrödinger cat state \cite{NoteEvenOdd} is given by
\begin{equation}\label{State:even}
    \begin{aligned}
    \ket{\phi_\text{even}}
    &= \dfrac{1}{\sqrt{N}}
        \big[
             \xi'_1 \ket{\Tilde{\alpha}(\theta) + \delta\alpha \cos \theta}\\
             &\hspace{1.7cm}
             + \xi_2 \ket{\Tilde\alpha(\theta) - \delta \alpha \cos\theta}
        \big],
    \end{aligned}
\end{equation}
where $\Tilde{\alpha} \equiv (\alpha-\delta\alpha)/\cos\theta$. As we can see, the distance between both coherent states is now given by $2\cos\theta$.

\begin{figure}
    \centering
    \includegraphics[width =0.48\textwidth]{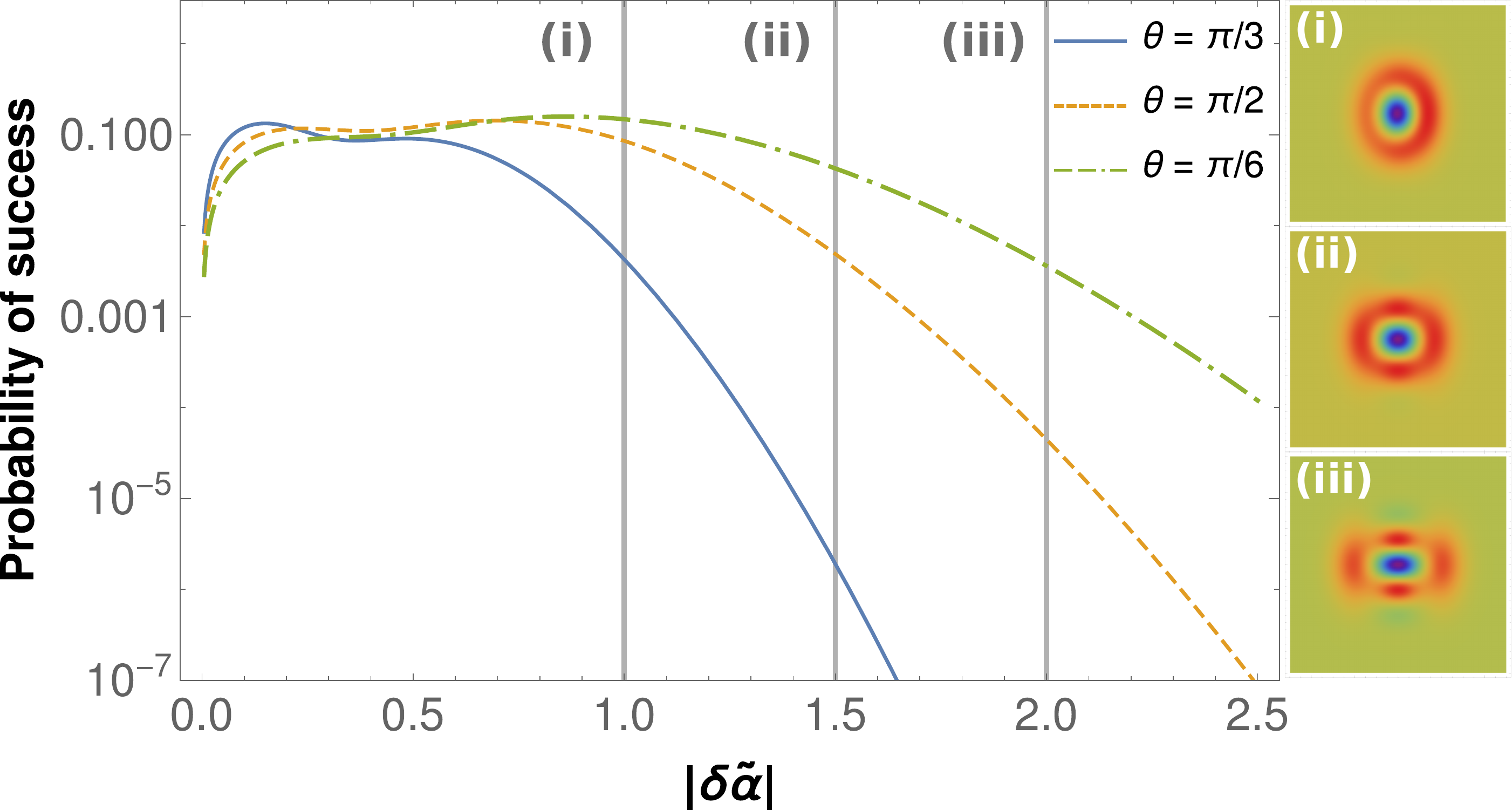}
    \caption{Probability of success when conditioning the results of the second mode to obtain an odd number of photons, versus $\delta \Tilde{\alpha} \equiv \delta \alpha \cos(\theta)$ (in absolute value). We present the results for three different beam splitters characterized by mixing angles $\theta = \pi/3$ (blue continuous line), $\theta = \pi/2$ (orange dashed line) and $\theta = \pi/6$ (green dashed-dotted line). The obtained results are similar to the ones obtained with the even number of photons measurement, but slightly enhanced. The vertical continuous grey lines correspond to the values of $|\delta \Tilde{\alpha}|$ for which we get the different Wigner functions shown in the insets (i)-(iii) on the right. In these plots we have set $\Tilde{\alpha} = 0$ so that the functions are centered in the origin.}
    \label{Fig8}
\end{figure}

One of the main concerns regarding these operations is the probability of being successful when performing the conditioning measurement over one of the beam splitter modes (see Appendix \ref{Appendix:A} for its analytic derivation). In Fig.~\ref{Fig7} we present this quantity for three beam splitters characterized by mixing angles $\theta = \pi/3$ (blue continuous line), $\theta = \pi/2$ (orange dashed line) and $\theta = \pi/6$ (green dashed-dotted line). The reflectivity/transmisivity ratio in these three cases are 75/25, 50/50 and 25/75, respectively. As $\delta \Tilde{\alpha} \equiv \delta \alpha \cos \theta$ increases, the success probability decays almost exponentially for the three angles. However, depending on the specific values the decay behaves differently. Of particular interest is the case of $\theta = \pi/6$, for which the distance between the coherent states is the largest and given by $\sqrt{3}\delta \alpha$. In this case, the maximum success probability is slightly smaller than 0.1, but the decay for increasing values of $\delta \Tilde{\alpha}$ takes place more slowly than in the other cases. In fact, for $|\delta \Tilde{\alpha}| \geq 1.5$, it clearly surpasses the other three cases, with a success probability on the order of $10^{-2}$ to $10^{-3}$ up to $|\delta \Tilde{\alpha}| = 2$. 

On the other hand, instead of conditioning the second mode over the measurement of an even number of photons, we can instead condition over an odd number. In that situation, the final state is similar to the one shown in Eq.~\eqref{State:even} but with a relative phase of $\pi$ between both states in the superposition, i.e. an \emph{odd} optical Schrödinger cat state. Furthermore, as can be seen in Fig.~\ref{Fig8}, this measurement allow us to slightly enhance the final probability of success.

In the insets located in the right part of Figs.~\ref{Fig7} and \ref{Fig8} we present the density plot of the measured Wigner functions in a displaced frame of reference for different values of $|\delta \Tilde{\alpha}|$, in particular $|\delta \Tilde{\alpha}| = 1.0, 1.5$ and $2.0$. They present the typical form of even and odd \emph{dead/alive} cat states: as $|\delta \Tilde{\alpha}|$ increases, the two coherent states in the superposition get far away one from the other, which leads to the appearance of two positive peaks in the Wigner function located in $x = \pm \delta \Tilde{\alpha}$, as can be seen clearly for the case $|\delta \Tilde{\alpha}| = 2.0$ (inset (iii) in Figs.~\ref{Fig7} and \ref{Fig8}). In such case, the maxima and minima appearing in between corresponds with their overlap, which are interchanged for even and odd cat states: the negative parts in this intermediate region appearing for the even cat state, turn into minima for the odd cat state and viceversa. Furthermore, in this case we have performed our calculations with the state in Eq.~\eqref{Enhanced:Superposition}, which means that for small values of $|\delta \Tilde{\alpha}|$ the final Wigner function is not completely symmetric as can be seen more clearly for the odd cat state at $|\delta \Tilde{\alpha}| = -1.0$ in Fig.~\ref{Fig8} (i), for which the positive ring around the negative minima is not symmetric.

\section{Conclusions}\label{Conclusions}
In this work we have studied how can we use the \emph{dead/alive} cat states generated with HHG in order to obtain superpositions containing three coherent states, and we have characterized them by looking at their Wigner function. While the first method is based on an interferometer approach, where in each arm we perform HHG, the second one is based on sequential HHG processes over the same mode. Here, we have analysed some particular configurations that allow us to obtain a wide variety of states, as we could see from the analysis of their Wigner functions which show different tendencies for increasing values of the generated shift $\delta \alpha$ depending on the specific set-up. However, there exists a large set of possibilities depending on where do we implement the conditioning measurements, and also on the different optical elements we can use in between.

Additionally, we demonstrated that we can use these three coherent state superpositions for the generation of enlarged versions of even and odd \emph{dead/alive} cat states, which could be very beneficial for practical purposes in quantum computation and quantum information. However, the main drawback of this method is that it relies on a heralding measurement over one of the outputs of a beam splitter, whose success probability decays almost exponentially as the desired \emph{dead/alive} cat state becomes larger. Regarding this, possible ways of improving the present results might be oriented towards the performance of more complex measurements over the second mode that can enhance the final probability distribution. Therefore, this motivates further investigation towards the generation of non-classical light states with HHG.

Finally, in this work we have highlighted the potential role of strong-field physics towards quantum optics. Given the wide variety of strong-field processes that one can generate in the laboratory apart from HHG, a natural question that arises is how their quantum analysis could be advantageous for practical purposes in fields like quantum computation and quantum information. Here we have indirectly tackled this question by altering and improving the HHG Schrödinger cat states. Nevertheless, other strong-field processes, like Above-Threshold Ionization where the electron recollides elastically with its parent ion \cite{Faria2020}, could lead to the desired superpositions more directly. Thus, the above discussion stimulates theoretical and experimental research towards the connection between strong-field physics and practical uses of quantum information processing.

%
%

\begin{acknowledgements}
ICFO group acknowledges support from ERC AdG NOQIA, Agencia Estatal de Investigaci\'on (``Severo Ochoa'' Center of Excellence CEX2019-000910-S, Plan National FIDEUA
PID2019-106901GB-I00/10.13039/501100011033, FPI), Fundació Privada Cellex, Fundació Mir-Puig, and from Generalitat de Catalunya (AGAUR Grant No. 2017 SGR 1341, CERCA program, QuantumCAT\_U16-011424, co-funded by ERDF Operational Program of Catalonia 2014-2020), MINECO-EU QUANTERA MAQS (funded by State Research Agency (AEI) PCI2019-111828-2/10.13039/501100011033), EU Horizon 2020 FET-OPEN OPTOLogic (Grant No 899794), and the National Science Centre, Poland-Symfonia Grant No. 2016/20/W/ST4/00314, Marie Sklodowska-Curie grant STRETCH No 101029393. J.R-D. acknowledges support from the Secretaria d'Universitats i Recerca del Departament d'Empresa i Coneixement de la Generalitat de Catalunya, as well as the European Social Fund (L'FSE inverteix en el teu futur)--FEDER. P.T. group acknowledges LASERLABEUROPE (H2020-EU.1.4.1.2 Grand ID 654148)), FORTH Synergy Grant AgiIDA (Grand No.: 00133). ELI-ALPS is supported by the European Union and co-financed by the European Regional Development Fund (GINOP Grant No. 2.3.6-15-2015-00001).
\end{acknowledgements}

\section*{Declarations}
\noindent\textbf{Conflicts of interest} The authors declare that they have no conflict of interest

\noindent\textbf{Code and data availability} The code used for the generation of the presented figures is made available in \cite{FigureMaker} under the Creative Commons Attribution-ShareAlike 4.0 (CC BY-SA 4.0) license.

%
%



\appendix
\section{Probability of success in the generalization of the conditioning approach}\label{Appendix:A}
Here we present a more detailed analysis about the probability of success plotted in Figs.~\ref{Fig7} and \ref{Fig8}. In the text we are interested in measuring even or odd number of photons in one of the output modes of the beam splitter, as this operation projects the other mode into a cat state, which differ in a relative phase of $\pi$ depending on the specific measurement that is performed. Thus, the set of positive operator-valued measurements characterizing the operations that we can implement is
\begin{equation}\label{POVM:large:cat}
    \bigg\{ \mathbbm{1} \otimes \dyad{0},
        \mathbbm{1} \otimes \sum_{n=1}^{\infty} \dyad{2n},
        \mathbbm{1} \otimes \sum_{n=0}^{\infty} \dyad{2n+1}
    \bigg\},
\end{equation}
which project the second mode into the vacuum state, an even number of photons and an odd number of photons respectively.

Each of the measurements described in Eq.~\eqref{POVM:large:cat} have a probability associated, so we define the \emph{probability of success} $\mathcal{P}$ as the probability of successfully performing one of such measurements. For instance, the probability of successfully measuring an even number of photons is given by
\begin{equation}
    \mathcal{P}_\text{even}
        = \dfrac{1}{\mathcal{P}_\text{total}} \tr{P_\text{even}\rho},
\end{equation}
where $\rho$ is the density matrix characterizing the state of our system, and $\mathcal{P}_\text{total}$ is given as the sum of all the possible measurements that we can perform, i.e.
\begin{equation}
    \mathcal{P}_\text{total}
    = \mathcal{P}_\text{zero}
    + \mathcal{P}_\text{even}
    + \mathcal{P}_\text{odd}.
\end{equation}

With all this set, let us consider the architecture presented in the main text used for the generation of enlarged cat states. Using a beam splitter characterized by the mixing angle $\theta$ and a coherent state of the form $\ket{(\alpha - \delta \alpha) \tan(\theta)}$, the final state after the beam splitter is given by
\begin{equation}\label{After:generic:BS}
    \begin{aligned}
    \ket{\Phi_\text{BS}}
        &= \dfrac{1}{\sqrt{N}}
            \Big[
                 - \xi_1' \ket{\Tilde{\alpha} 
                    + \delta \alpha \cos\theta} \ket{-\delta \alpha \sin\theta}\\
                 &\hspace{1.1cm}
                 + \ket{\Tilde{\alpha}}\ket{0}
                 - \xi_2 \ket{\Tilde{\alpha} 
                 - \delta \alpha \cos\theta} \ket{\delta \alpha \sin\theta}
             \Big],
    \end{aligned}
\end{equation}
where $\Tilde{\alpha} = (\alpha-\delta\alpha)/\cos\theta$. Thus, the measurement of an even number of photons projects the state in Eq.~\eqref{After:generic:BS} into
\begin{equation}
    \begin{aligned}
    \ket{\Phi_\text{even}}
        &= \dfrac{1}{\sqrt{N}}
            \Big[
                 -\xi_1' \ket{\Tilde{\alpha}+\delta \alpha \cos\theta}
                 - \xi_2  \ket{\Tilde{\alpha}-\delta \alpha \cos\theta}
            \Big]\\
        &\hspace{0.5cm}
            \otimes
            \sum_{n=1}^\infty \dfrac{(\delta \alpha \sin\theta)^{2n}}{\sqrt{2n!}} e^{-\tfrac{|\delta \alpha \sin \theta|^2}{2}} \ket{2n}\\
        &\equiv \ket{\phi_\text{even}}
            \otimes
            \sum_{n=1}^\infty \dfrac{(\delta \alpha \sin\theta)^{2n}}{\sqrt{2n!}} e^{-\tfrac{|\delta \alpha \sin \theta|^2}{2}} \ket{2n};
    \end{aligned}
\end{equation}
the measurement of an odd number of photons projects the state into
\begin{equation}
    \begin{aligned}
    \ket{\Phi_\text{odd}}
        &= \dfrac{1}{\sqrt{N}}
            \Big[
                 \xi_1' \ket{\Tilde{\alpha}+\delta \alpha \cos\theta}
                 - \xi_2  \ket{\Tilde{\alpha}-\delta \alpha \cos\theta}
            \Big]\\
        &\hspace{0.5cm}
            \otimes
            \sum_{n=0}^\infty \dfrac{(\delta \alpha \sin\theta)^{2n+1}}{\sqrt{2n+1!}} e^{-\tfrac{|\delta \alpha \sin \theta|^2}{2}} \ket{2n+1}
            \\
        &\equiv \ket{\phi_\text{odd}}
            \otimes
            \sum_{n=0}^\infty \dfrac{(\delta \alpha \sin\theta)^{2n+1}}{\sqrt{2n+1!}} e^{-\tfrac{|\delta \alpha \sin \theta|^2}{2}} \ket{2n+1};
    \end{aligned}
\end{equation}
and, finally, the measurement of zero number of photons leads to
\begin{equation}
    \begin{aligned}
    \ket{\Phi_\text{zero}}
        &= \dfrac{1}{\sqrt{N}}
          \Big[
                -\xi_1' \braket{0}{-\delta \alpha \sin\theta}
                \ket{\Tilde{\alpha} + \delta \alpha \cos \theta}
                + \ket{\Tilde{\alpha}}\\
                & \hspace{1.1cm}
                - \xi_2 \braket{0}{\delta \alpha \sin\theta} \ket{\Tilde{\alpha} - \delta \alpha \cos \theta}
          \Big]\otimes \ket{0}\\
        &\equiv
        \ket{\phi_\text{zero}} \otimes \ket{0}.
    \end{aligned}
\end{equation}

Thus, the probability of success for each of the three measurements that we can do is given by
\begin{equation}
    \mathcal{P}_\text{even}
        = \dfrac{\braket{\phi_\text{even}}}{\mathcal{P}_\text{total}}
            e^{-|\delta \alpha \sin\theta|^2}
            \big[
                 -1 + \cosh(\delta\alpha \sin\theta)
            \big],
\end{equation}
\begin{equation}
    \mathcal{P}_\text{odd}
        =
        \dfrac{\braket{\phi_\text{odd}}}{\mathcal{P}_\text{total}}
            e^{-|\delta \alpha \sin\theta|^2}
            \sinh(\delta\alpha \sin\theta),
\end{equation}
and
\begin{equation}
    \mathcal{P}_\text{zero}
        =
        \dfrac{\braket{\phi_\text{zero}}}{\mathcal{P}_\text{total}}.
\end{equation}

Note that the case of $\delta \alpha = 0$ leads to a divergence in the obtained equations. In that case the applied measurement does not make sense at all because the case of $\delta \alpha = 0$ corresponds to the situation in which we do not have HHG processes taking place. Thus, no optical Schrödinger cat state is being generated.
\end{document}